\documentclass[a4paper,usenatbib]{mnras}
\usepackage[T1]{fontenc}
\usepackage{ae,aecompl}
\usepackage{graphicx}
\usepackage{hyperref}
\usepackage{amsmath}
\usepackage{amssymb}
\usepackage{mathtools}
\usepackage{bm}
\usepackage{cleveref}

\title[Starquakes and gravitational waves] 
{Starquakes in millisecond pulsars and gravitational waves emission} 

\author[Giliberti et al.]{
E. Giliberti$^{1,3}$\thanks{elia.giliberti@unimi.it},
G. Cambiotti$^{2}$ 
\\
$^{1}$Dipartimento di Fisica, Universit\`a degli Studi di Milano, Via Celoria 16, 20133,Milano, Italy\\
$^{2}$Dipartimento di Scienze della Terra, Universit\`a degli Studi di Milano, Via Cicognara 7, Milano, 20129, Italy\\
$^{3}$Istituto Nazionale di Fisica Nucleare, sezione di Milano, Via Celoria 16, 20133 Milano, Italy
}

\date{Accepted XXX. Received YYY; in original form ZZZ}

\pubyear{2020}

\begin{document}
\label{firstpage}
\pagerange{\pageref{firstpage}--\pageref{lastpage}}
\maketitle

\begin{abstract}
So far, only transient Gravitational waves (GWs) produced by catastrophic events of extra-galactic origin have been detected. However, it is generally believed that there should be also continuous sources of GWs within our galaxy, such as accreting neutron stars (NSs). In fact, in accreting NSs, centrifugal forces can be so strong to break the neutron star crust (causing a starquake), thus producing a quadrupole moment responsible for the continuous emission of GWs. At equilibrium, the angular momentum gained by accretion and lost via GWs emission should balance each other, stopping the stellar spin-up. 

We hereinafter investigate the above physical picture within the framework of a Newtonian model describing compressible, non-magnetized and self-gravitating NSs. In particular, we calculate the rotational frequency need to break the stellar crust of an accreting pulsar and we estimate the upper limit for the ellipticity due to this event. Depending on the equation of state (EoS) and on the mass of the star, we calculated that the starquake-induced ellipticity ranges from $10^{-9}$ to $10^{-5}$. The corresponding equilibrium frequency that we find is in good agreement with observations and, for all the scenarios, it is below the observational limit frequency of $716.36$ Hz. Finally, we also discuss possible observational constraints on the ellipticity upper limit of accreting pulsars.

\end{abstract}


\section{Introduction}

Gravitational waves (GWs) detections have widen our knowledge of astrophysical events. First the discover of black holes merger \citep{ligo_first_bh}, and then the neutron stars (NSs) coalescence \citep{ligo_first_ns} have opened new possible windows for the study of extreme compact objects. However, until now direct detections of GWs came only  from catastrophic, transient events, with an extra-galactic origin. 
We expect, nonetheless, that continuous signals should come also from our Galaxy, emitted by fast rotating, accreting pulsars.

Observation of Low-Mass X-ray Binaries (LMXBs) has shown a paucity of stars rotating near the centrifugal break-up frequency\footnote{This very rough estimate of the frequency beyond which the centrifugal force tears apart the star is given by the Keplerian rotational frequency $\nu_{limit}$=$\sqrt{G M/4 \pi a^{3}}$, where $M\simeq1.4 M_{odot},a\simeq10 km$ are the canonical stellar mass and radius, respectively, and $G$ is the gravitational constant.} \citep{lattimer2007,chakra2008}, opening the outstanding question of why these objects seems to spin well under that limit. In fact, preliminary estimates by \citet{cook1994} suggested that the spin-up timescale for NSs in LMXBs should be large enough to make them reach at least the rotational frequency of 1 kHz. On the contrary, the actual fastest spinning accreting NS has $\nu=599$ Hz \citep{galloway2005}, and also millisecond pulsars, that are thought to be the ultimate fate of LMXBs \citep{Bhattamillisecond}, rotate with a frequency lower than the break-up frequency. In particular, \citet{chakra2003} using Bayesian statistics have shown that the actual distribution of the Accreting Millisecond X-ray Pulsars (AMXPs) gives a theoretical maximum spin frequency of $760$ Hz.

One possible explanation for this behaviour is that these kind of pulsars emit GWs that make them slow down.
In particular, many works \citep{bildsten1998, usho2000, watts2008} suggest that accreting millisecond pulsars can reach an equilibrium configuration when the angular momentum gained from infalling material is lost by GWs emission. \citet{pringle78} and \citet{wagoner1984} firstly suggested that GWs emission can explain the dynamical equilibrium of NSs, but only more recently \citet{bildsten1998} proposed the possibility of mountains forming on accreting objects as a concrete mechanism for generating a non-zero ellipticity and, as a consequence, GWs.   

It is also interesting to remind that multi-million molecular dynamic simulations have shown that the crustal breaking strain can be quite large \citep{horo2009,baiko18} and, therefore, that the crust can sustain a maximum ellipticity large enough to generate GWs detectable from Earth by the current generation of interferometers \citep{haskell2006, owen2013}. 

In the literature, two main mechanisms able to produce a static ellipticity have been studied: thermal mountains \citep{bildsten1998, usho2000, haskell2015}, and magnetically confined mountains \citep{cutler2002, melatos2005, haskell2008, vigelius2009, priymak2011}. The former are due to pycno-nuclear reactions that heat the accreted material deep into the crust. The latter, instead, are caused by a local enhancements of the magnetic field structure (related to accretion) that can sustain mountains.


Using the higher breaking strain threshold, and modeling a NS as a homogeneous, incompressible object, \citet{fattoyev2018} claimed that starquakes (\citet{ruderman1969}, \citet{baym1969}) can happen only on accreting, rapidly rotating star, where the centrifugal force is large enough to make the crust reach the failure threshold. They also introduced the hypothesis that the breaking of the crust might produce a quadrupolar deformation sufficient to emit enough energy through GWs to prevent the stellar spin-up. However, in \citet{fattoyev2018} the ellipticity produced by starquakes is not self-consistently calculated as well as the resultant evolutionary path of the NS, i.e. the reaching of the equilibrium angular velocity. 

In this work we use the model described in \citet{giliberti_main} to study the deformation of a rotating, compressible, non-magnetized, self-gravitating NS to explore the idea that a sequence of starquakes can act as a trigger for GWs emission. 

In particular, we will study the following physical picture. A NS is accreting mass from a companion: the infalling material creates a disk that transfers angular momentum to the star and spins-up the central object. The NS will thus rotate faster and faster til the breaking condition is reached: in that moment a first starquake occurs, altering the stellar axial symmetry by creating a non-null ellipticity. From then on, the star radiates GWs. The balance between the angular momentum gained from accretion and the one lost by emission will bring the star through a \emph{sequence} of breaks and finally to a dynamical equilibrium frequency.

The paper's organization reflects the different steps needed to explore the NS's evolutionary path: first of all in Section \ref{sezione configurazione} we summarize the model described in \citet{giliberti_main}, focusing on the stellar configuration; in section \ref{sezione frequenza rottura} we study the problem of crust failure and find an estimation of what are the typical frequencies necessary to break the crust. Once it is shown that fast rotating stars can reach the breakup frequency, we move forward in section \ref{sezione ellitticita} by introducing an upper limit for the ellipticity $\epsilon$ due to a series of starquakes on the NS. In this way, in section \ref{sezione frequenza equilibrio} we will be ready to study the dynamical equilibrium frequency. 
Finally, in section \ref{sezione dei beta} we make a comparison of the ellipticity upper limit predicted by our model with the one estimated using observations of fast rotating pulsars.  

\section{Neutron star configuration}\label{sezione configurazione}
\begin{figure}
\includegraphics[width=\columnwidth]{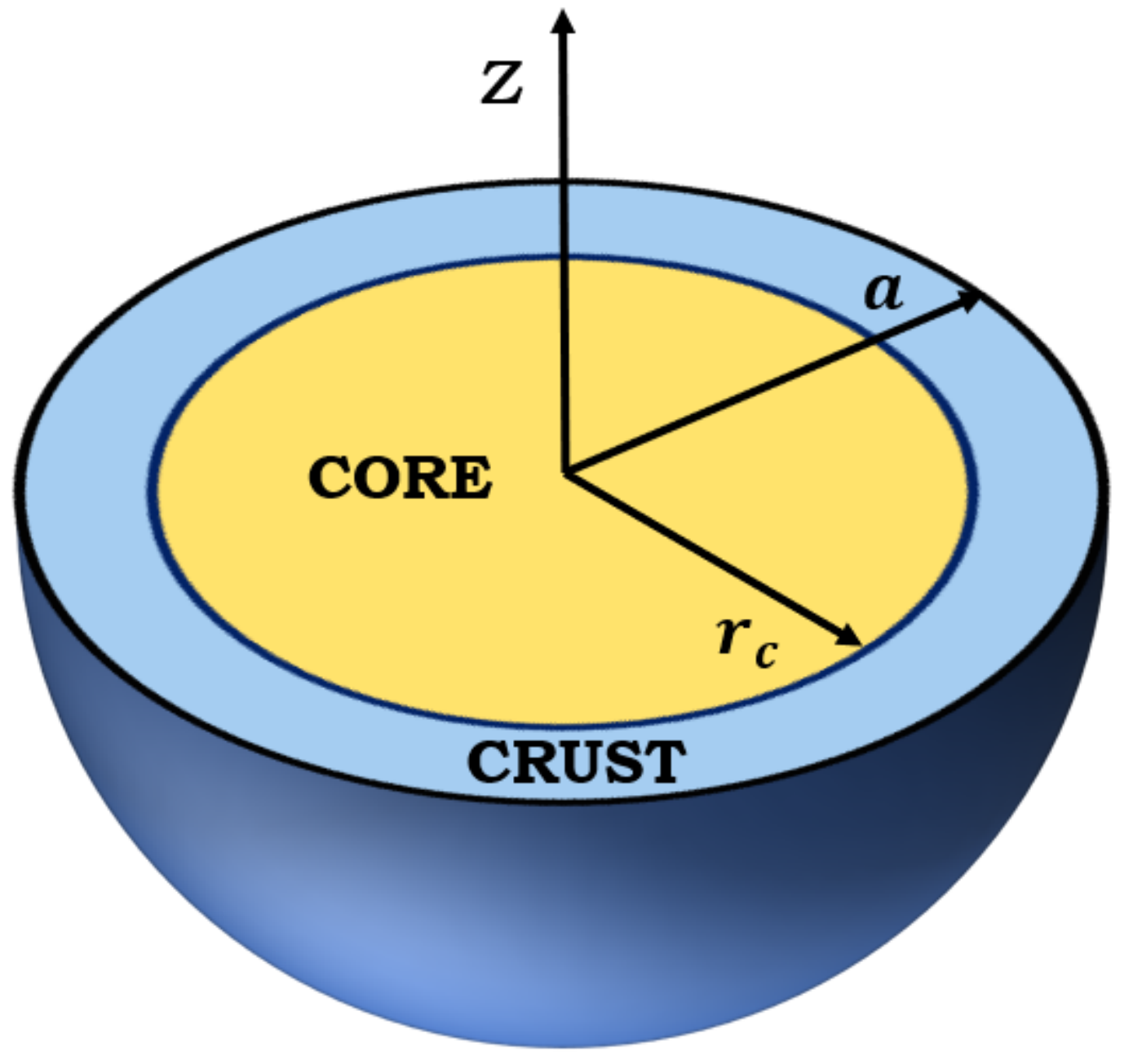}
\caption{Sketch of the stellar structure for the model considered in this work (not in scale). The star is divided into a fluid core, extending from the centre to $r=r_{c}$, and an elastic crust, that goes from the core-crust boundary up to the stellar surface $r=a$.}
\label{figura struttura stella}
\end{figure}

We want to analyze the stressing effect of material accreting from a companion on a NS, accelerating the object.
Our aim is to study the deformation of a self-gravitating, non-magnetized\footnote{In this work we are interested mainly on LMXB and millisecond pulsars, that have both typically very low magnetic field at the surface $B\approx10^{8-9}$ G \citep{CatalogoATNF}: we expect that in this condition $B$ has a very small impact on crustal deformation \citep{franco2000}.}, compressible NS, under the effect of the centrifugal force, in order to find what is the maximum rotational frequency before crust-breaking. For this purpose, we use a general Newtonian model \citep{giliberti_main}, where the star consists of a fluid core, extending from the origin to the radius $r_{c}$, and an elastic crust, that covers the region from $r_c$ to the stellar surface $r=a$\footnote{In a spherical coordinate system $r$ is the radial distance from the star's center, $\theta$ is the colatitude and $\varphi$ is the longitude.} (Fig. \ref{figura struttura stella}). The outer-crust boundary is placed at the density $1\times10^{11}\text{g/cm}^3$ in order to guarantee the numerical stability of the solution against the computational problems due to the very rapid variation of the density in the outermost layers \citep{usho2000}.
Following \citet{usho2000}, the crust-core transition is set at the fiducial density $1.5\times10^{14}\text{g/cm}^3 $, that implies a core-crust transition at $r_{c} \approx 0.90 \, a$ for a standard neutron star with $M=1.4M_{\odot}$.
We use MATHEMATICA 11 to perform numerical computation.

For describing the stellar matter, we choose the equation of state (EoS) of a polytrope of index $n=1$, since it allows us to study in a somewhat realistic way the star's physical characteristic as a function of its mass (for a fixed mass, a  Newtonian approach gives larger stars with respect to Relativistic ones). With this choice, the stellar mass and radius are independent of each other, giving us the opportunity to associate the Relativistic radius $a$ with a given mass $M$. In other words, we can use the realistic mass-radius relation of a chosen EoS, obtained from the integration of the Tolman-Oppenheimer-Volkoff equations, for fixing the star's radius $a$ once the stellar mass $M$ is chosen. In particular, in this paper we will use two different EoSs and their mass-radius relation: SLy \citep{douchin2001} and the stiffer BSk21 \citep{goriely2010}. Both are \emph{unified} EoSs, covering consistently both the core and the crust; furthermore, the SLy EoS is a standard choice for making predictions about crustal quadrupoles, see e.g. \citep{Horo10, owen2013}. The comparison between EoSs is extremely important, since it makes possible to see the impact of different stars configuration on the deformation of NSs.

Furthermore, following the analysis of \citet{giliberti_main}, we study the behaviour of the same star in two scenarios, parametrized  by different values of the adiabatic index. In the first, the response of the stellar matter to perturbations is characterized by its \emph{equilibrium} adiabatic index, namely $\gamma^*=2$. This is the situation in which the typical dynamical timescale are small compared to the reactions ones. In the second scenario, instead, the contrary is true and the adiabatic index differ from its equilibrium value. We refer to it as the frozen adiabatic index $\gamma_f$ and consider the two cases of $\gamma_f=2.1$ and $\gamma_f=\infty$. We choose this two values to give a preliminary insight of the spectrum of all the possible allowed frozen indices: $\gamma_f=2.1$ mimics a small departure from the equilibrium case, while $\gamma_f=\infty$ represents an incompressible response of an initial compressible star to external forces. We refer to \citet{giliberti_main} for a deep discussion on the difference between these two scenarios and their impact on displacements and stresses of the star's crust.

In our model, once the EoS and the value of the adiabatic index have been chosen, there are only other two parameters needed to complete the description of the NS's configuration. They are the stellar elastic moduli: the bulk modulus $\kappa$ and the shear modulus $\mu$. The first is given by the relation
\begin{equation}
    \kappa(r)=\gamma\,P(r),
\end{equation}
where $P$ is the local pressure. In particular, the initial non rotating configuration will be always characterized by $\kappa=\gamma^*P$, while the response of the star to the centrifugal force will be modeled with $\kappa=\gamma^*P$ or $\kappa=\gamma_{f}P$, according to the dynamical timescale of the spin-up.
Concerning the shear modulus we use the same prescription as \citet{cutler2003}
\begin{equation}
    \mu=10^{-2} P.
    \label{shear modulus}
\end{equation}

\section{Breaking frequency}\label{sezione frequenza rottura}

Starting from an unstressed, spherical, symmetric, non-rotating configuration\footnote{We are interested in calculating the displacement field between a configuration rotating with velocity $\Omega$ and one rotating at $\Omega+\delta\Omega$, where $\delta\Omega>0$ for a spinning up pulsar. The non-rotating configuration is known for our elastic star, since it coincides with the one given by the usual hydrostatic equilibrium for a fluid. Thanks to the assumed linearity of the problem, the calculated displacement will be proportional to the difference $(\Omega+\delta\Omega)^2-\Omega^2$.}, the spin-up caused by the infalling material coming by the companion will cause a deformation of the star, that will become oblate. The stress in the crust will grow as the rotational velocity increases, till a breaking condition is reached. We use the Tresca criterion to establish when the crust will break, i.e. when we have a starquake. This criterion states that the crust will fail when the \emph{strain angle} $\alpha$, defined as the difference between the maximum and the minimum eigenvalues of the strain tensor, is half of the breaking strain $\sigma_{max}$ \citep{failure_book}
\begin{equation}
    \alpha=\frac{\sigma_{max}}{2}.
    \label{Tresca}
\end{equation}
%
Since the deformation due to rotation is proportional to the frequency squared
\begin{equation}
    \alpha=\Tilde{\alpha}\nu^2,
    \label{alpha di nu}
\end{equation}
the Tresca criterion will be satisfied at a threshold frequency, that we call \emph{breaking frequency} $\nu_{b}$. $\Tilde{\alpha}$ is a term depending only on the structure of the NS, namely its mass and EoS. From Eqs \eqref{Tresca} and \eqref{alpha di nu} we immediately get
\begin{equation}
    \nu_{b}=\sqrt{\frac{\sigma_{max}}{2\Tilde{\alpha}_{max}}},
    \label{breaking freq}
\end{equation}
where $\Tilde{\alpha}_{max}$ is the maximum values of $\Tilde{\alpha}(r,\theta)$ on the whole NS crust, i.e. in the range $r_{c}\leq r\leq a$, $0\leq \theta \leq \pi$ ($\theta$ is the colatitude angle). Therefore, the larger $\sigma_{max}$, the larger the breaking frequency, as expected.

Unfortunately the value of $\sigma_{max}$ is very uncertain, ranging from $10^{-5}$ of the first theoretical estimation of \citet{rudermanII1991}, to the more recent values obtained with molecular dynamic simulations of $10^{-1}$ by \citet{horo2009} or of $0.04$ proposed with semi-analytical approaches by \citet{baiko18}. 
In the present work we use the larger breaking strain $10^{-1}$ for two reasons. The first is to compare our results with the ones obtained by \citet{fattoyev2018}, that used the same threshold. The second is to give an upper limit for the breaking frequency $\nu_{b}$. In fact, in the case of the lowest estimation of $\sigma_{max}=10^{-5}$, we can see (cf Eq \eqref{breaking freq}) that the breaking frequency will decrease of about two orders of magnitude compared to our choice.

\begin{figure}
\includegraphics[width=\columnwidth]{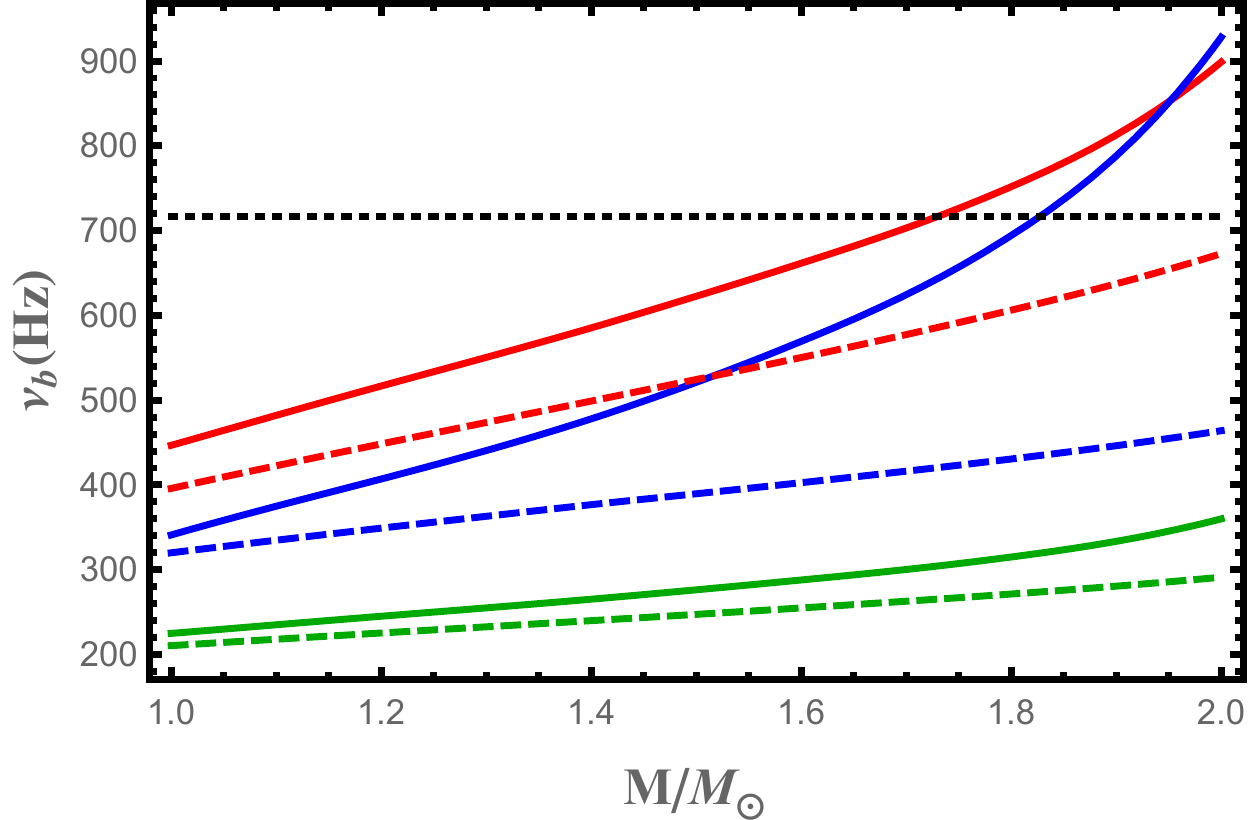}
\caption{The breaking frequency $\nu_{b}$ as a function of the stellar mass for different EoS and adiabatic indices. The solid lines are obtained with the SLy EoS, while the dashed ones for the BSk21 one. Different colours indicates different adiabatic indices values: red for $\gamma^*=2$, blue for $\gamma_{f}=2.1$ and green for $\gamma_{f}=\infty$. The black, dotted line indicates the rotation frequency of the fastest rotating pulsar observed} $\nu_{o}=716.36$ Hz \citep{Hessels2006}.
\label{figura frequenza rottura}
\end{figure}

The results of this first analysis, coming from the model briefly introduced in section \ref{sezione configurazione}, are shown in Fig \ref{figura frequenza rottura}, where the curves for $\nu_b$ are plotted for different EoSs and different values of the adiabatic index. The main features of Fig \ref{figura frequenza rottura} are the following:
\begin{enumerate}
    \item 
    For a given NS mass, a softer EoS produces less compact stars and, so, gives larger breaking frequency values. Indeed, as shown by \citet{giliberti_pasa} and \citet{giliberti_main}, the star's compactness, $M/a$, is a key parameters controlling the deformations.

    \item The larger the adiabatic index value, the lower the breaking frequency. 
    We know \citep{giliberti_main} that incompressible star ($\gamma_{f}=\infty$) will develop larger strains $\Tilde{\alpha}_{max}$ with respect to a compressible one: using Eq \eqref{breaking freq} it is clear that this means that the larger $\gamma$ the smaller the breaking frequency.
    
    \item A small change in the adiabatic index value gives large changes in the breaking frequency curve. 
    As an example, for the BSk21 EoS, we can see that the $\nu_{b}$ curve for $\gamma_f=2.1$ lies exactly in the middle between the one for $\gamma^*=2$ and the other for $\gamma_{f}=\infty$. This is a typical features of compressible, self-gravitating NSs, due to the smallness of the shear modulus with respect to the bulk modulus \citep{chamel_livingreview}. In fact, if the ratio $\mu/\kappa$ is small, the star, despite its elastic crust, behaves essentially like a fluid. 
    This means that the incompressible limit is reach even with a small departure of the adiabatic index from its equilibrium \citep{giliberti_main}.

    \item For a typical $M=1.4M_{\odot}$ NS, we can say that the breaking frequency is in the range $200-600$ Hz, well below the maximum observed rotational frequency \citep{Hessels2006}
    \begin{equation}
    \nu_{o}=716.36\,\mathrm{Hz}.
    \label{freq osservativa max}
    \end{equation}
    In this respect, our analysis refines the results obtained by \citet{fattoyev2018}, which predict larger breaking frequencies in the range $400-1000$ Hz.
\end{enumerate}

Fig \ref{figura frequenza rottura} deserves a last comment. There are some stars which have a breaking frequency larger than $\nu_{o}$. The reasons why so far we have not observed any NS with this frequency could be many and different, and are all compatible and understandable within our model. Indeed we expect very few NSs rotate with a frequency larger than $\nu_{o}$. In fact, the majority of NSs show a breaking frequency that is smaller than $700$ Hz and only
massive stars with a softer EoS can reach a breaking frequency of $800-900$ Hz.

Moreover, we can think that the fast NSs we can observe are actually near their \emph{equilibrium frequency}, i.e. the frequency at which the angular momentum gained from accretion is equal to the one lost by GWs emission, that must not be confused with the \emph{breaking frequency}, which is the typical frequency at which the crust starts to fail. And, as we will see in the following sections, the equilibrium frequency is typically smaller than the breaking one.

\section{Starquakes induced ellipticity}\label{sezione ellitticita}
\begin{figure}
\includegraphics[width=\columnwidth]{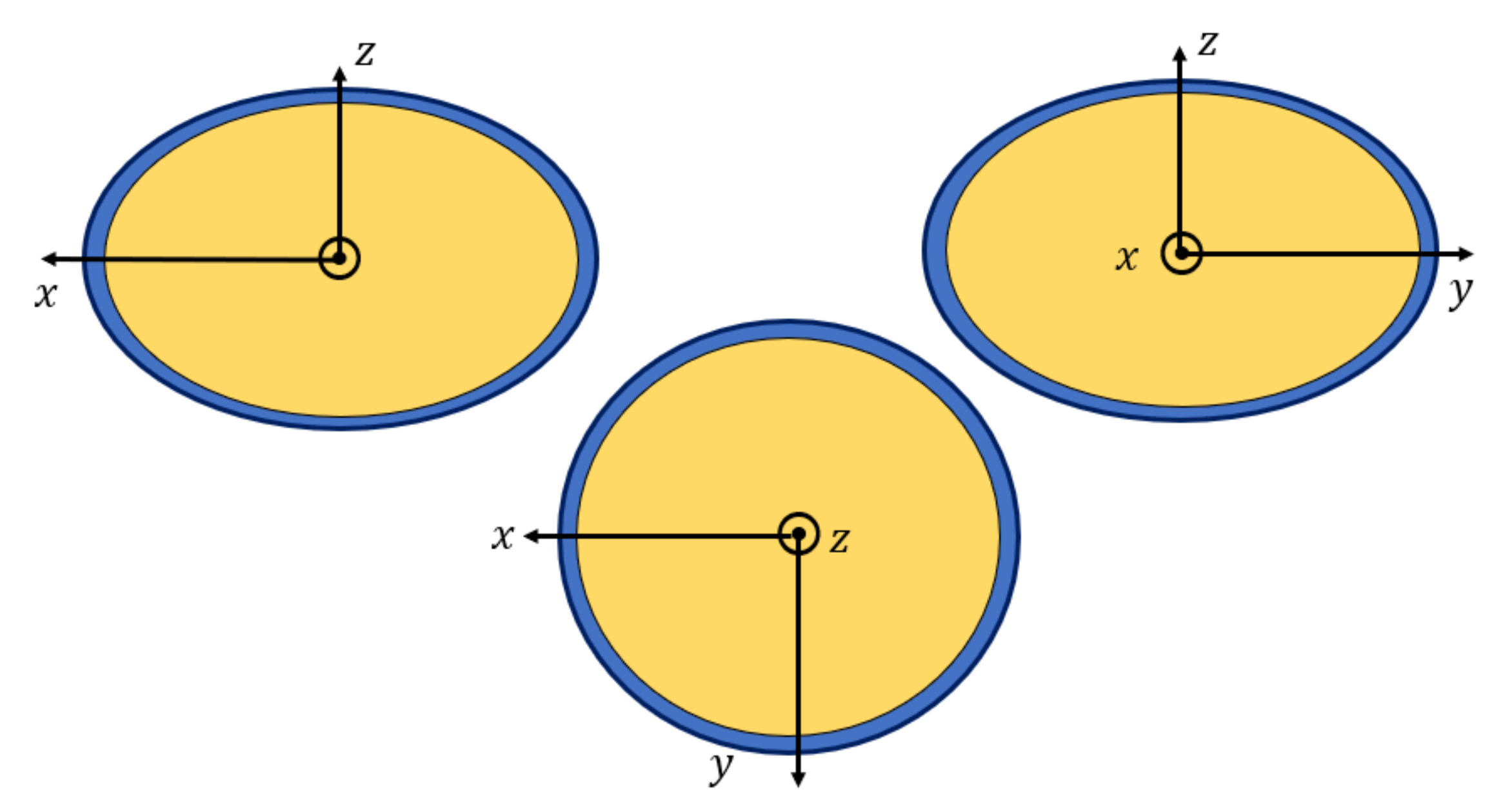}
\includegraphics[width=\columnwidth]{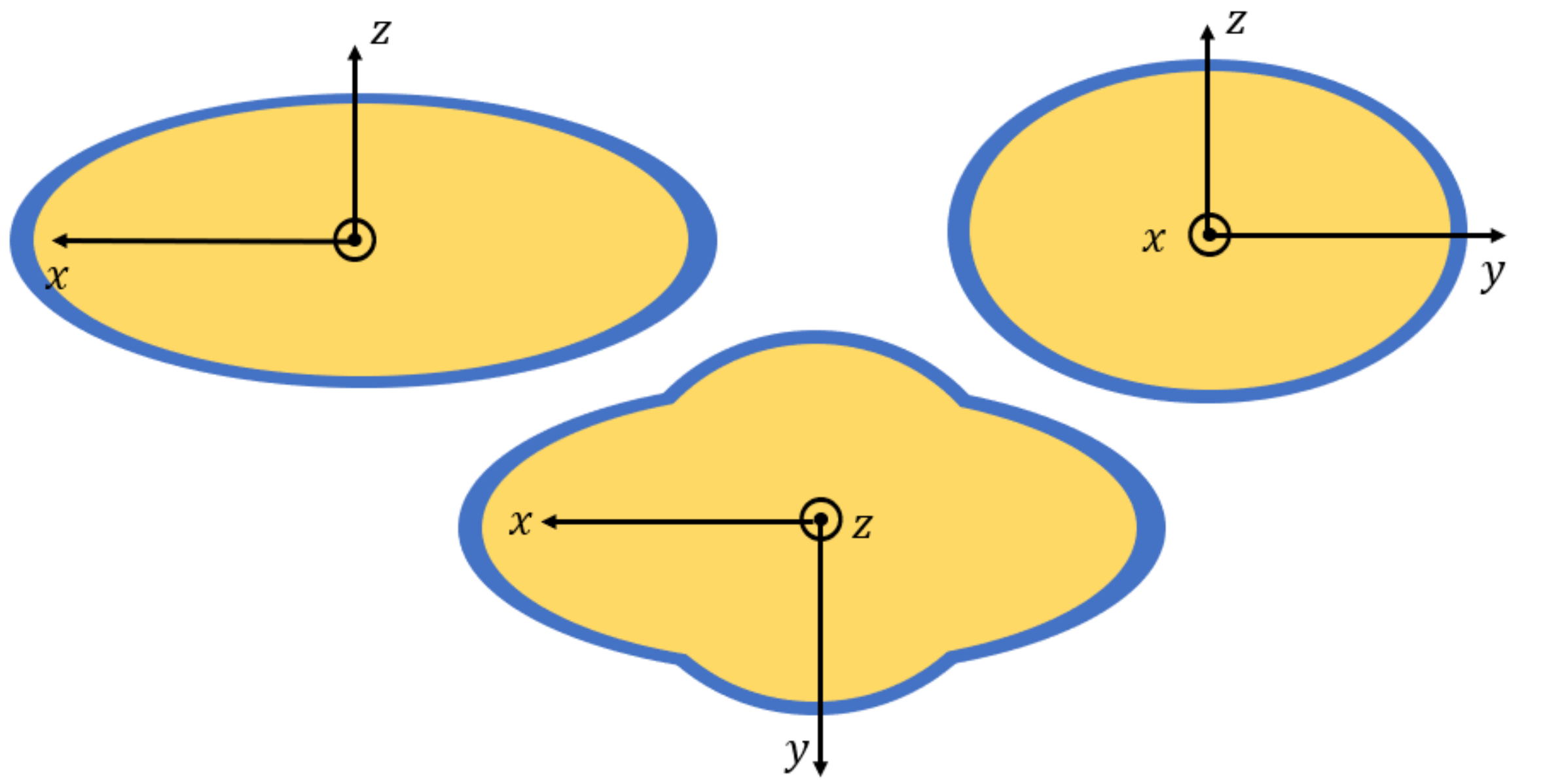}
\caption{Sketch of the stellar deformation (not in scale). The star is divided into a fluid core (yellow) and an elastic crust (blue).
\emph{Top}: Vision of the star from the three axis $x, y, z$ in the uniform rotation configuration.
\emph{Bottom}: Vision of the star from the three principal axis $x, y, z$ after a starquake that has created a mountain in the $x$ direction.}
\label{figura deformazione stella}
\end{figure}

Now that we know that the sufficiently fast rotating NSs can reach the condition for crust breaks, we can study the possible consequence of starquake events and, as a consequence, the creation of a non-null ellipticity. In fact, a break of the crust cause a local deformation, that brings the star away from the pure axyal simmetry, see Fig. \ref{figura deformazione stella}. This deformation can be evaluated by the ellipticity $\epsilon$, defined as
\begin{equation}
    \epsilon=\frac{A-B}{C},
    \label{ellitticità}
\end{equation}
where $A$ and $B$ are the principal moments of inertia along the equatorial axes, while $C$ is the one along the rotation axis. 
The exact value of $\epsilon$ depends on the crustal properties of the star and on its seismic history. As a first approach, one would be led to estimate the effect of every single starquake; however this is a too difficult task.  In fact, our knowledge of the NS crust physics is extremely poor, and a reasonable description of a quake involves a very large number of unknown parameters (dip and strike angles, displacement discontinuity, fault area etc.). For these reason it seems more reasonable to follow a different approach.

We look at starquakes as the attempts of a stressed star to achieve the equilibrium fluid shape despite the constraining action of its elastic crust. As Eq \eqref{ellitticità} shows, in the simple case of an uniform rotation the star is axially symmetric, i.e. $A=B$ and thus $\epsilon=0$. Whatever the angular velocity, the crust of the star is stressed, since it cannot achieve the corresponding equilibrium configuration that it \emph{would have if it was completely fluid}. The elastic crust, in fact, constrains the star to have a more prolate shape with respect to the fluid one. 
However, the axial symmetry of centrifugal deformation can be broken by starquakes, that can create a \emph{mountain} on the NS surface. Starting from an initial (pre-starquake) configuration, through a sufficient number of breaking events, the star will, therefore, tend towards its fluid configuration.  In other words, the cumulative effect of a sequence of many crust failures is to give a more oblate shape to the star. Clearly, it sound physically reasonable to state that the dynamical equilibrium will be achieved with a \emph{sequence} of events, since we do not expect that a single quake could release all the stresses of the crust and bring instantaneously the star to its fluid configuration. 
On the other hand, the breaking of the crust leads the NS to get an ellipticity different from zero and thus, to emit GWs. 

Thus, in order to calculate the maximum ellipticity (i.e. the one reached after a ``complete" sequence of starquakes) that a NS could have at a given angular velocity, we compare the principal moment of inertia of two different configurations, rotating at the same frequency. The first is the one of a NS with a solid crust, while the second is the one of a pure fluid star. The difference between these two configuration, in terms of moment of inertia, will give us the maximum value of $\epsilon$.

 
\subsection{Inertia tensor}
Let us introduce how to use our model for calculating the NS's inertia tensor. In the following, we adopt the notation of \citet{sabadini_book}; for a brief summary of the notation used, main quantities and equations introduced, see Appendix  \ref{appendice a}.

Consider an initially non-rotating star that is spun up to a given frequency $\nu$. The centrifugal perturbation acting on the object involves both the $\ell=0$ and $\ell=2$ spherical harmonics; however, since we are interested into the calculation of the stellar ellipticity, we can focus only on the latter. 
In fact, the $\ell=0$ term would give the same contributions to all the principal moments of inertia and these contributions cancel each other in the difference $A-B$. Furthermore, as far as the difference $A-B$ is small, the contribution to $C$ can be neglected within a first-order perturbation theory.
Therefore, at the first order approximation Eq \eqref{ellitticità} can be written as: 
\begin{equation}
    \epsilon=\frac{\Delta A- \Delta B}{I_0},
\end{equation}
where $\Delta A$ and $\Delta B$ are the changes of the principal moments of inertia $A$ and $B$ due to $\ell=2$ spherical harmonic perturbations, and $I_0$ is the unstressed stellar moment of inertia.

In the Cartesian reference frame, the changes of the inertia tensor $I_{ij}$ due to $\ell=2$ spherical harmonic perturbation can be obtained according to the following expression
\begin{equation}
\Delta I_{ij}=\sum_{m=-2}^{2}Q^{2m}_{ij} \frac{4\pi}{5}\int_{0}^{a}\rho_{2m}^{\Delta}\left(r\right)r^{4}dr,
\label{inerzia l=2}
\end{equation}
where $\rho_{2m}^{\Delta}$ and $Q^{2m}_{ij}$ are, respectively, the spherical harmonic coefficients of degree $\ell=2$ and order $m$ of the density distribution and of the matrix $Q_{ij}$ defined by
\begin{equation}
    Q_{ij}=\frac{1}{3}\delta_{ij}-\hat{r}_{i}\hat{r}_{j}.
    \label{definizione Q}
\end{equation}
Here $\delta_{ij}$ is the Kronecker delta and $\hat{r}_i$ are the Cartesian components of the radial unit vector.
The deformation of the star is described by two Poisson equations, one for the perturbed gravitational potential $\phi^{\Delta}$
\begin{equation}
    \nabla^2\phi^{\Delta}=4\pi G \rho^{\Delta},
\end{equation}
and the other for the centrifugal potential $\phi^{C}$
\begin{equation}
    \phi^{C}=-8\pi^2 \nu^2.
\end{equation}
It is quite natural to introduce the total perturbed potential $\Phi^{\Delta}$, as the sum of $\phi^{\Delta}$ and $\phi^{C}$.
By expanding also the total perturbed potential in spherical harmonics we can write \citep{cha01987} %
\begin{equation}
\Phi_{2m}^{\Delta}\left(a\right)=-\frac{4\pi G}{5a^{3}}\int_{0}^{a}\rho_{2m}^{\Delta}\left(r\right)r^{4}dr.
\label{Relation phi rho}
\end{equation}
Therefore, from Eqs \eqref{inerzia l=2} and \eqref{Relation phi rho}, it follows that the perturbed 
tensor of inertia can be written as: 
\begin{equation}
\Delta\boldsymbol{I}_{ij}=-\sum_{m=-2}^{2}Q^{2m}_{ij}\frac{a^{3}}{G}\Phi_{2m}^{\Delta}\left(a\right).
\label{I2 in funzione del potenziale}
\end{equation}
The above expression is extremely useful for the calculation of the inertia changes, since it involves only the value of the total potential at the star's surface, that can be easily obtained within the model presented in \citet{giliberti_main}.
 We also observe that in the case of a uniform rotation the inertia tensor can be expressed in a diagonal form, namely 
\begin{equation}
    \Delta\boldsymbol{I}=\mathrm{Diag}[\Delta A, \Delta A,\Delta C],
    \label{relazione inerzia uniforme}
\end{equation}
where $\Delta C$ is the change of the moments of inertia along the rotational axis $C$.
Finally, we note that  by choosing a coordinate system in which the rotational axis $z$ coincides with $\theta=0$, we can restrict ourselves to the only 
$\ell=2$, $m=0$ spherical harmonic, and write
\begin{equation}
    \boldsymbol{Q}^{20}=Diag[1/3,1/3,2/3].
    \label{equazione Q}
\end{equation}
In this case $\Delta A$ and $\Delta C$ satisfy the relation
\begin{equation}
    \Delta C=-2\Delta A=-\frac{2}{3}\frac{a^{3}}{G}\Phi_{20}^{\Delta}\left(a\right).
    \label{traccia nulla A C}
\end{equation}
In fact, the perturbation terms due to the $\ell=2,m=0$ spherical harmonic contributes only to the deviatoric part of the inertia tensor.

With the Equations \eqref{I2 in funzione del potenziale}-\eqref{traccia nulla A C} we are now equipped to deduce an upper limit for the ellipticity caused by starquakes, as it will be shown in the following subsection.

\subsection{Estimation of ellipticity}

%
%


The pre-starquakes (rotating, stressed, elastic) and final (rotating, fluid) configurations will be characterized by two slightly different inertia tensors, that can both be written in a diagonal form, $\Delta\boldsymbol{I}^{E}=\mathrm{Diag}[\Delta A^{E}, \Delta A^{E},\Delta C^{E}]$ and $\Delta\boldsymbol{I}^{F}=\mathrm{Diag}[\Delta A^{F}, \Delta A^{F},\Delta C^{F}]$, where $E$ stands for \emph{elastic} and $F$ for \emph{fluid}, respectively.
The explicit calculation of these two tensors is done by using Eqs \eqref{I2 in funzione del potenziale}, \eqref{relazione inerzia uniforme} and \eqref{equazione Q}.

If we assume, as said above, that a pure fluid star will be more oblate with respect to an elastic one at the same rate of rotation, we can state that
\begin{align}
    \Delta C^{E}\leq \Delta C^{F}
    \label{relazioni C fluidi e solidi}
\\   
\Delta A^{F}\leq \Delta A^{E}.
    \label{relazione A fluidi e solidi}
\end{align}
%
In between the initial and the final fluid-like configurations, also the tensor of inertia - that is not necessarily symmetric due to the intrinsic nature of the rupture process - can be given in the diagonal form. Considering again only the deviatoric $\ell=2$ harmonic term, we can therefore write, using the apex $Q$ for the post-quakes configuration:
\begin{equation}
    \boldsymbol{I}^{Q}=\mathrm{Diag}[\Delta A^{Q},\Delta B^{Q},\Delta C^{Q}],
\end{equation}
where $\Delta B$ is the change of the moments of inertia along one of the principal axis. In this case $\Delta A^{Q}\neq \Delta B^{Q}$.
Following above considerations, we expect that the post-quakes configuration will be ``something" between the elastic and the fluid ones; therefore we require that
\begin{align}
   \Delta C^{E}\leq \Delta C^{Q} \leq \Delta C^{F}\nonumber\\
    \Delta A^{F}\leq \Delta A^{Q},\Delta B^{Q} \leq \Delta A^{E}.
    \label{relazioni tra i coefficienti}
\end{align}
%
%
%
Using Eq \eqref{relazioni tra i coefficienti} we see that the maximum difference $\Delta B^{Q}-\Delta A^{Q}$ can be expressed as $\Delta A^{E}-\Delta A^{F}$, and thus we are able to obtain an upper limit for the ellipticity due to a series of starquakes (see Eq \ref{traccia nulla A C}):
\begin{equation}
    \epsilon_{max}=\frac{a^{3}}{3 I_0 G}\left[\Phi_{20}^{\Delta F}\left(a\right)-\Phi_{20}^{\Delta E}\left(a\right)\right].
    \label{epsilon max}
\end{equation}
Therefore, to compute $\epsilon_{max}$, we have to build two different rotating configurations for the same star: the first has an elastic crust, as sketched in Fig \ref{figura struttura stella}, while in the second the object is completely fluid.\footnote{As explained briefly in section \ref{sezione configurazione}, our configuration is fixed with the choice of the NS's mass, EoS and adiabatic index value.} For each of these two configurations we can extract the perturbed total potential value at the stellar surface, calculate the corresponding change in the inertia tensor through Eq \eqref{I2 in funzione del potenziale} and, finally, get the value of the maximum ellipticity using Eq \eqref{epsilon max}. 
In order to model the the elastic configuration, we used a typical shear modulus for cold catalyzed matter (Eq \eqref{shear modulus}) since the fact that accreting NSs can reach very high temperature in the crust ($10^8$ K) should not affect much the shear modulus shape \citep{chamel_livingreview}.
Moreover, high temperature can lower the $\mu$ value, which means that the crust is more similar to a fluid than  it is in the cold configuration \citep{Hoffman2012}. Since we are looking for for the \emph{maximum} difference between the elastic and fluid configuration, our choice 
follows straightforwardly.

In Table \ref{tab epsilon max} are reported the values of $\epsilon_{max}$ for a $M=1.4M_{\odot}$ NS rotating at $\nu=\nu_{o}$ (Eq \eqref{freq osservativa max}) both with SLy and BSk21 EoSs. As we can see, also in this case a small change in the value of the adiabatic index value causes a large difference in the ellipticity's value. This results can be compared with the ones of \citet{owen2013}, which are typically used in the GWs literature as benchmark. Note that if the star is in its equilibrium configuration (i.e. $\gamma^*=2$), the upper limit value that we find is lower than the maximum value of ellipticity ($\epsilon\approx10^{-5}$) that a standard  $1.4 M_{\odot}$ NS can sustain \citep{owen2013}. But, as soon as we depart from this condition and $\gamma=2.1$ or more, our model predicts upper limit values for ellipticity even exceeding $10^{-5}$. 
However, we underline that these two estimated values of ellipticities comes from very different perspective. In fact, the one of \citet{owen2013} is the maximum elastic deformation that the star can sustain \emph{before} breaking, while the ellipticity given by our Eq \eqref{epsilon max} is the upper limit of the deformation that can be reached \emph{due to} the breaking process.
Therefore, from our model we expect that starquakes could produce somewhat large ellipticities in accreting (or fast rotating) NSs and in turn GWs.

Our model shows also a strong dependence of the star's ellipticity on the stellar mass, see Table \ref{tab rapporto epsilon}. For both the EoSs a $M=1M_{\odot}$ object can produce an ellipticity about one order of magnitude larger than an heavier $M=2M_{\odot}$ star. 

%
\begin{table}
\centering{}
\begin{tabular}{|r|c|c|c|}
& $\gamma^*=2$ & $\gamma_f=2.1$ & $\gamma_f=\infty$ \\ 
\hline
SLy &$5.2\times10^{-7}$&$1.5\times10^{-5}$&$2.5\times10^{-5}$ \\ 
\hline
BSk21  &$1.3\times10^{-6}$&$3.3\times10^{-5}$&$5.5\times10^{-5}$ \\ 
\hline
\end{tabular}
\caption{
Maximum ellipticity \eqref{epsilon max} calculated with SLy and BSk21 EoSs for a standard NS with mass $M=1.4M_{\odot}$. In all the cases $\nu=\nu_{o}$ (Eq \eqref{freq osservativa max}).
}
\label{tab epsilon max}
\end{table}
%
\begin{table}
\centering{}
\begin{tabular}{|r|c|c|c|}
& $\gamma^*=2$ & $\gamma_f=2.1$ & $\gamma_f=\infty$ \\ 
\hline
SLy &$39$&$30$&$25$ \\ 
\hline
BSk21  &$11$&$8$&$6$ \\ 
\hline
\end{tabular}
\caption{
Ratio of the maximum ellipticity of Eq \eqref{epsilon max} for a $M=1 M_{\odot}$ NS and $M=2M_{\odot}$ NS, i.e. $\frac{\epsilon_{max}(1M_{\odot})}{\epsilon_{max}(2M_{\odot})}$, calculated with SLy and BSk21 EoSs.
}
\label{tab rapporto epsilon}
\end{table}


\section{Equilibrium frequency}\label{sezione frequenza equilibrio}
In section \ref{sezione frequenza rottura} we have shown that typical $1.4M_{\odot}$ NSs rotating with frequencies in the range $200- 700$ Hz may undergo a series of starquakes, and consequently, emit GWs. 
Now we will focus our attention on the consequences that this emission might have on the NSs' dynamical equilibrium.

Our NS is tearing some material from its companion and so it is gaining angular momentum, with a rate $N_{acc}$ that is roughly given by \citep{usho2000}
\begin{equation}
N_{acc}=\dot{M}\sqrt{GMa}.
\label{accrescimento}
\end{equation}
At the same time the star, that reached the crust breaking condition and that has a non-null ellipticity, is losing angular momentum through GWs emission at a rate $N_{GW}$. This latter can be written, using standard symbols, as \citep{usho2000} 
\begin{equation}
N_{GW}=\frac{128\pi^{3}}{5}\frac{GI^{2}\nu^{5}\epsilon^{2}}{c^{5}}.
\label{gravitational waves}
\end{equation}
%

We can now use 
$\epsilon_{max}$ to get a \emph{lower} limit for the equilibrium frequency value. In fact, since the crust ruptures are caused by fast rotation, we can express $\epsilon_{max}$ as
\begin{equation}
    \epsilon_{max}=\Tilde{\epsilon}\nu^2,
    \label{definizione epsilon}
\end{equation}
where $\Tilde{\epsilon}$ is a function of the EoS and of the stellar mass.
By equating Eq \eqref{accrescimento} and Eq \eqref{gravitational waves}, and using the above expression \eqref{definizione epsilon} we get the \emph{equilibrium frequency}
\begin{equation}
    \nu_{eq}=K\frac{\dot{M}^{1/9}(Ma)^{1/18}}{I^{2/9}\Tilde{\epsilon}^{2/9}},
    \label{nu equilibrio epsilon max}
\end{equation}
where we have explicitly written all the terms depending on the stellar mass and EoS, and $C$ is the constant defined by
\begin{equation}
    K=\frac{\sqrt[9]{\frac{5}{2}} c^{5/9}}{2 \pi ^{5/9} \sqrt[18]{G}}.
\end{equation}
From Eq \eqref{nu equilibrio epsilon max} we can obtain $\nu_{eq}$, both for SLy and BSk21 EoSs, as a function of the mass and of the adiabatic index value. 
The dynamical equilibrium clearly depends on the rate $\dot{M}$ at which the star is accreting. We consider two different thresholds that roughly constraints the region where the astrophysical values for this kind of objects can be found. In particular, they are the same values given by \citet{usho2000}: an upper limit of $\dot{M}=2\times10^{-8}M_{\odot}/\mathrm{yr}$ and a lower one of $\dot{M}=10^{-10}M_{\odot}/\mathrm{yr}$. 

\begin{figure}
\includegraphics[width=\columnwidth]{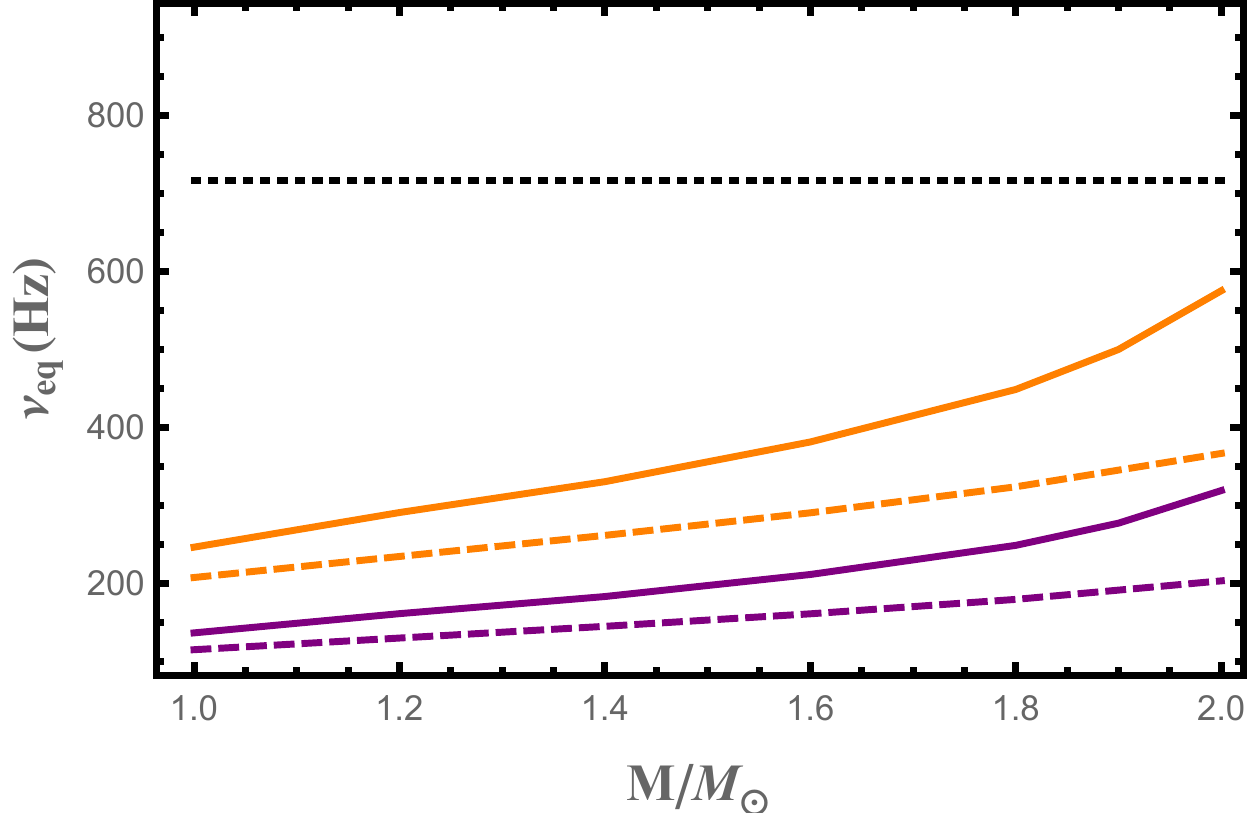}
\caption{Equilibrium frequency $\nu_{eq}$ as function of the stellar mass for SLy (solid) and BSk21 (dashed) EoSs, fixed $\gamma^*=2$. The curves are calculated for two different mass accretion rates: $\dot{M}=2\times10^{-8}M_{\odot}/\mathrm{yr}$ (orange) and $\dot{M}=10^{-10}M_{\odot}/\mathrm{yr}$ (purple). The black, dashed line represents the actual maximum observed rotational frequency $\nu_{o}$.}
\label{frequenza equilibrio gamma due}
\end{figure}

The results for $\gamma^*=2$ are shown in Fig \ref{frequenza equilibrio gamma due}, while the study of the effect of different adiabatic indeces on the equilibrium frequency is exemplified for a $M=1.4M_{\odot}$ NS in Tables \ref{tab SLy} and \ref{tab BSk21}. In fact, we expect that a more compressible star has a smaller maximum ellipticity and thus a larger equilibrium frequency, if compared with an incompressible one, and this is exactly what happens. 
The value of $\gamma^*=2$ in Fig \ref{frequenza equilibrio gamma due} has been chosen since the curves plotted in this case are the largest ones between the equilibrium and the frozen scenario. 

Let us note that, as said in section \ref{sezione frequenza rottura}, the expected equilibrium frequency is smaller than $\nu_o$. Furthermore, $\nu_{eq}$ (Tables \ref{tab SLy} and \ref{tab BSk21}) is always lower than the breaking frequency (cf. Fig \ref{figura frequenza rottura} with Fig \ref{frequenza equilibrio gamma due}), i.e.
\begin{equation}
\nu_{eq}<\nu_{b}.    
\label{freq equilibrio e breaking}
\end{equation}
Despite the different values obtained for different adiabatic indices, the above relation remains valid. 
This relation suggest the following physical picture. An old star accretes some material from a companion, increasing its angular velocity. Stresses develop into the crust, till the breaking strain is reached: crust fails, the star loses its axial symmetry and starts to emit GWs. The rate of angular momentum lost by this emission is greater than the one gained from accretion, and the star spins-down, till the equilibrium is reached.
However, we remind that the estimated values of $\nu_{eq}$ calculated in this way are the \emph{lower} value, since in Eq \eqref{nu equilibrio epsilon max} we have used our \emph{upper} limit for the ellipticity. Furthermore, since the breaking frequency is calculated for $\sigma_{max}=10^{-1}$, 
when the breaking strain is smaller, we get a lower breaking frequency.

Thus, from our analysis we expect the breaking of the crust for rapidly rotating pulsars, even in the case of very high ($10^{-1}$) breaking strain; moreover, our model, through the development of a large ellipticity, suggests also why we don't observe any NS with rotational frequency above $700$ Hz.
\begin{table}
\centering{}
\begin{tabular}{|r|c|c|c|}
\hline
&$\gamma^*=2$ & $\gamma_f=2.1$ & $\gamma_f=\infty$\\ 
\hline
$\nu_{b}$ (Hz) &585 &478 &265\\ 
\hline
$\nu_{eq}^{Max}$ (Hz)  &331 &157 &140\\ 
\hline
$\nu_{eq}^{Min}$ (Hz)  &183 &87 &78\\ 
\hline
\end{tabular}\caption{
Breaking frequency ($\nu_{b}$) and equilibrium frequencies ($\nu_{eq}$) calculated with SLy EoS for a $M=1.4M_{\odot}$ NS and different mass accretion rates: $\dot{M}=2 \times10^{-8}\dot{M_{\odot}}/\mathrm{yr}$ ($\nu_{eq}^{Max}$) and $\dot{M}=1 \times10^{-10}\dot{M_{\odot}}/\mathrm{yr}$ ($\nu_{eq}^{min}$).
}
\label{tab SLy}
\end{table}
\begin{table}
\centering{}
\begin{tabular}{|r|c|c|c|}
& $\gamma^*=2$ & $\gamma_f=2.1$ & $\gamma_f=\infty$ \\ 
\hline
$\nu_{b}$ (Hz) &499 &377 &240 \\ 
\hline
$\nu_{eq}^{Max}$ (Hz)  &262 &128 &114 \\ 
\hline
$\nu_{eq}^{Min}$ (Hz)  &145 &80 &63 \\
\hline
\end{tabular}
\caption{
Breaking frequency ($\nu_{b}$) and equilibrium frequencies ($\nu_{eq}$) calculated with BSk21 EoS for a $M=1.4M_{\odot}$ NS and different mass accretion rates: $\dot{M}=2 \times10^{-8}\dot{M_{\odot}}/\mathrm{yr}$ ($\nu_{eq}^{Max}$) and $\dot{M}=1 \times10^{-10}\dot{M_{\odot}}/\mathrm{yr}$ ($\nu_{eq}^{min}$).
}
\label{tab BSk21}
\end{table}
%


\section{Observational constraints on ellipticity}\label{sezione dei beta}

In the previous sections, we used our upper limit value $\epsilon_{max}$ 
to calculate the NSs' equilibrium rotational frequency. However, just why it is an \emph{upper limit}, we do not expect all the sources to reach the maximum possible deformation measured by $\epsilon$. In this section we will follow a somewhat reverse path. Starting from observational data coming from electromagnetic and GWs observations we will constraint the ellipticity of observed NSs.

\subsection{Constraints from GWs non-detection}

The $O1$, $O2$ and (partially) $O3$ runs of LIGO/Virgo detectors has been used to search GWs coming from rapidly rotating NS. Searches focused both on wide-parameter sources \citep{Abbott1,Abbott2,Abbott3,Abbott4,Abbott5,Abbott6,Abbott7} and signal coming from specific target \citep{AbbottS1,AbbottS2,AbbottS3}. In particular, the recent paper \citep{AbbottS3} put constraints on the fiducial ellipticity for a selection of rapidly rotating ($\nu_r>100$ Hz) pulsars. These estimations can be very useful if compared with our maximum ellipticity value, Eq \eqref{epsilon max}: in fact, we can assume that during their life real pulsars reach only a fraction $\beta\leq1$ of our threshold, i.e.
\begin{equation}
    \epsilon=\beta\epsilon_{max}.
\end{equation}
If we state that the ellipticity of NSs is due \emph{only} to the starquakes mechanism, we can extract the value of $\beta$ simply as the ratio between LIGO/Virgo fiducial ellipticities $\epsilon_{L/V}$ and our upper limit 
\begin{equation}
    \beta_{GW}=\frac{\epsilon_{L/V}}{\epsilon_{max}}.
    \label{def beta gravitazionale}
\end{equation}
In this way, $\beta_{GW}>1$ means that the observations are still not constraining enough the $\epsilon$ value; if, on the contrary, $\beta_{GW}<1$ we are measuring how large is the fraction of $\epsilon_{max}$ currently developed on the NS.

\begin{figure}\centering
\includegraphics[width=0.75\hsize]{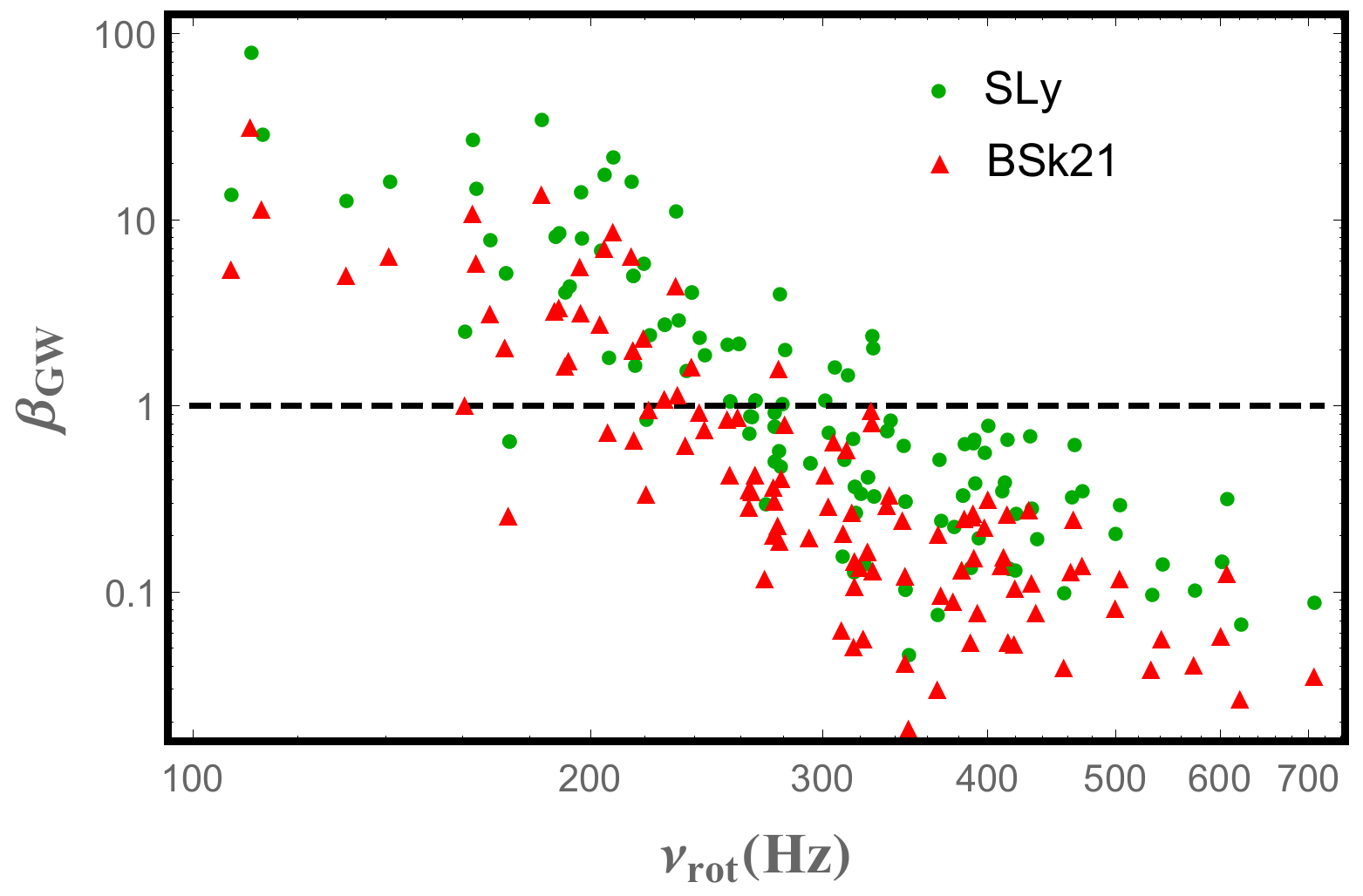}
\caption{Ratio of the observational ellipticity given by \emph{non-detections} of continuous GWs and our upper limit $\epsilon_{max}$. $\beta_{GW}$ \eqref{def beta gravitazionale} is calculated for $M=1.4 M_{\odot}$ with two different EoS: SLy (green circles) and BSk21 (red triangles). The dashed, black curve indicates $\beta_{GW}=1$.}
\label{BETA per run 2}
\end{figure}

In Fig \ref{BETA per run 2} we show $\beta_{GW}$, obtained using the definition \eqref{def beta gravitazionale} and calculated for $1.4M_{\odot}$, with both the EoSs. Using the SLy EoS we get a larger value of $\beta_{GW}=0.047$, while for BSk21 $\beta_{GW}=0.019$. However, these limit as to be kept with caution since, as have seen in section 4.2 both NS's mass and adiabatic index value can change these estimation of about one order of magnitude. 
Note that for the slowest stars we have always $\beta_{GW}>1$, but estimations for pulsars close to Earth could be relevant in the next future. For example, the recent constraints \citep{AbbottIsolate} for J043-4715 ($\epsilon<9.5\times10^{-9}$) and J071-6830 ($\epsilon<7.7\times10^{-9}$) are comparable with our upper limit of about $\epsilon_{max}\simeq10^{-8}$ for both pulsars.  
The latest LIGO-Virgo observational run (O3) is very important: the complete analysis of its signal, integrated over many months, could show the first direct detection of continuous GWs (and thus a measurement of $\epsilon$); if this were not the case, we could use the new data to lower the estimated value of $\epsilon$ (i.e., the estimation of $\beta$). In the debate whether to search for GWs emitted by slowly rotating or fast rotating objects our model suggests to search GWs emission from high rotating pulsars that we expect to have larger ellipticities (and therefore greater gravitational emission power) than the slowly rotating ones. 

\subsection{LMXB}
Observation of LMXBs can also be used to extract a value of $\epsilon$, giving an useful benchmark to compare it with our upper limit $\epsilon_{max}$. In fact, assuming that \emph{the measured rotational frequency of an observed LMXB is its equilibrium frequency}, one can obtain the corresponding ellipticity 
\begin{equation}
    \epsilon_{acc}=\frac{C^{9/2}}{\nu_{r}^{9/2}}\frac{\left\langle \dot{M}\right\rangle ^{1/2}\left(Ma\right)^{1/4}}{I},
    \label{nu equilibrio epsilon max beta}
\end{equation}
where \emph{acc} stands for \emph{accretion} and $\left\langle \dot{M}\right\rangle$ is the average mass accretion rate \emph{during outburst}\footnote{Typically accreting NSs show short bursts, lasting from days to months, with a corresponding high accretion rate,  and very long period of recovery, during which the accretion is orders of magnitude smaller than in active phase \citep{watts2008}.}.
In the following, we use the data elaborated by \citet{haskell2015} (see Table 1 therein), that give, for each star, its rotational frequency, distance and mass accretion rate. 
It is convenient to introduce the parameter 
\begin{equation}
    \beta_{acc}=\frac{\epsilon_{acc}}{\epsilon_{max}}.
    \label{definizione beta LMXB}
\end{equation}
The meaning of $\beta_{acc}$ is straightforward: every star with $\beta_{acc}\leq1$ has an equilibrium ellipticity that can be explained by the starquakes mechanism alone; on the contrary, for the ones with $\beta_{acc}>1$ we need to invoke some other effect (deformations due to a non-zero magnetic field, for example) to explain the equilibrium ellipticity.
In our calculations, the stellar mass is fixed at $M=1.4M_{\odot}$ while the adiabatic index value at $\gamma^*=2$: the values of $\epsilon_{max}$ for $\gamma_{f}=2.1, \infty$ are always larger than the one obtained with $\gamma^*$, giving a lower $\beta_{acc}$.

\begin{figure}
\includegraphics[width=\columnwidth]{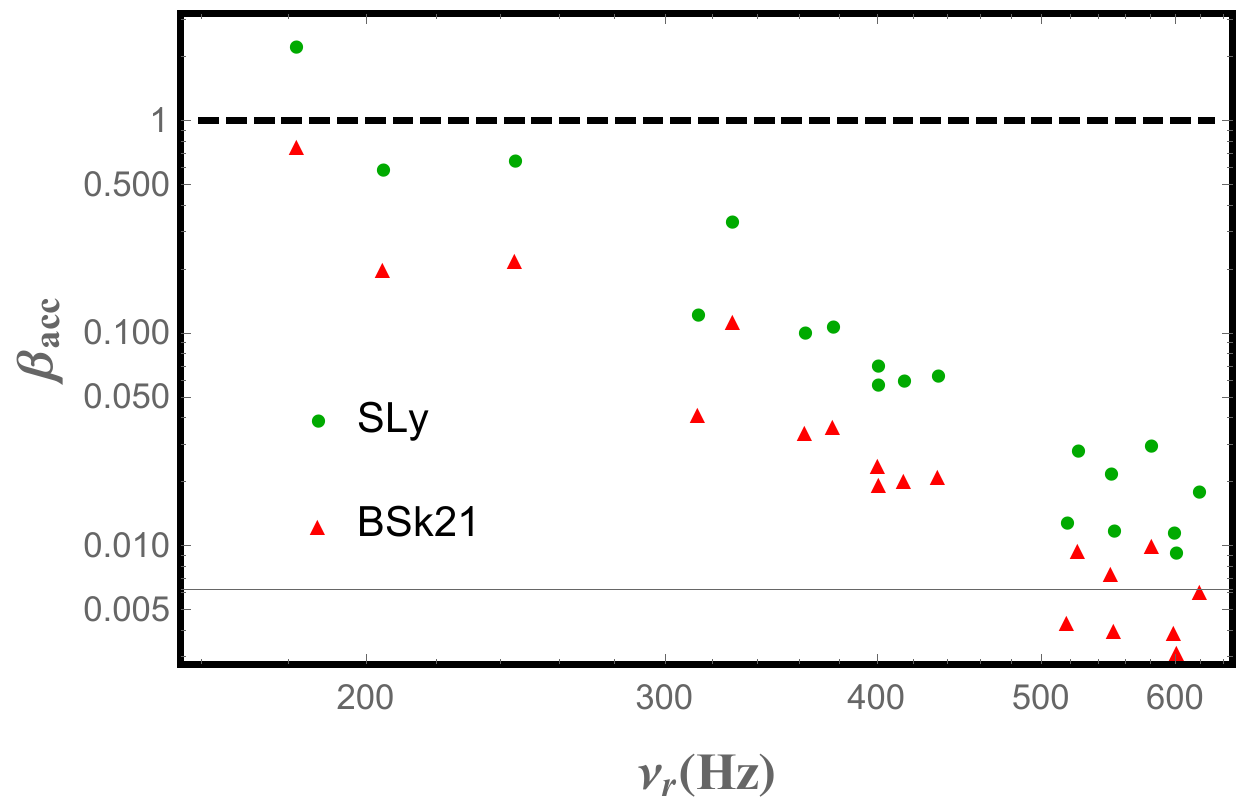}
\caption{Ratio of the observational ellipticity assuming dynamical equilibrium in LMXB objects $\epsilon_{acc}$ (Eq \eqref{definizione beta LMXB}) and our upper limit $\epsilon_{max}$ (Eq \eqref{epsilon max}) for a $M=1.4M_{\odot}$ NS with $\gamma^*=2$. Green dots are calculated with SLy EoS while red triangles with BSk21. The dashed, black curve indicates $\beta_{acc}=1$.}
\label{BETA equilibrio osservativo vs max}
\end{figure}

The value of $\beta_{acc}$ as a function of observed rotational frequency is shown shown in Fig \ref{BETA equilibrio osservativo vs max}.
%
%
About $95\%$ of the stars fall in the category $\beta_{acc}\leq1$ both using BSk21 or SLy EoS. The smallest value of $\beta_{acc}$ found with this method is $0.003$ for BSk21 and $0.09$ for SLy EoS.
In our sample only J1756.9-2508, in the case of Sly EoS, has a value of $\beta_{acc}$ larger than 1; this is due to the fact that this star has a rotating frequency of ``only" $182$ Hz, and thus has an $\epsilon_{max}$ value smaller than that of other stars. These results confirms that starquakes mechanism could explain why these stars have all a frequency smaller than $700$ Hz. In fact, even a small fraction of our maximum value $\epsilon_{ max}$ is enough for LMXBs to reach a dynamical equilibrium at frequency smaller than $\nu_o$.

\subsubsection{Alternative estimation of $\beta$}

\begin{figure}
\includegraphics[width=\columnwidth]{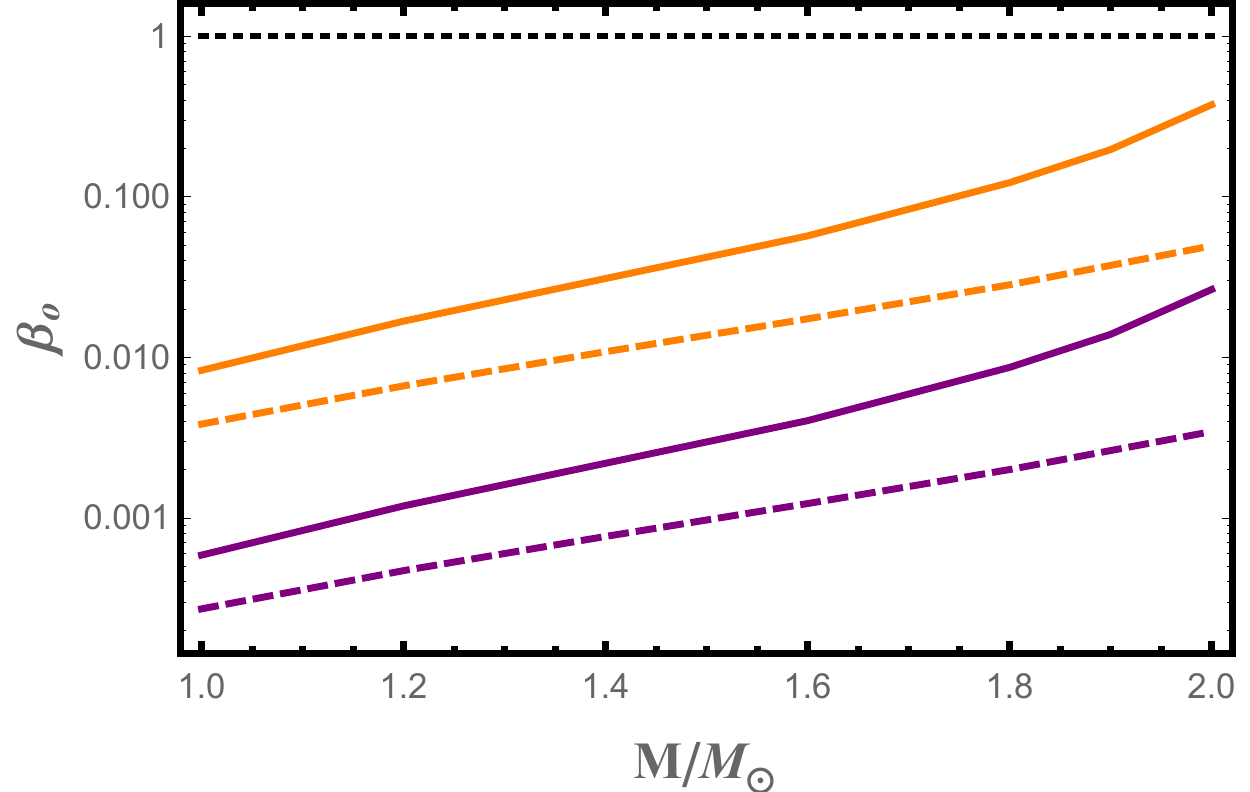}
\caption{Minimum value of $\beta_{o}$, defined implicitly in Eq \eqref{beta di m}, for SLy (solid) and BSk21 (dashed) EoSs. The curves are calculated for a NS with a mass of $M=1.4 M_{\odot}$, $\gamma^*=2$ for two different mass accretion rates: $\dot{M}=2\times10^{-8}M_{\odot}/\mathrm{yr}$ (orange) and $\dot{M}=10^{-10}M_{\odot}/\mathrm{yr}$ (purple). 
The black, dotted line represents $\beta=1$.}
\label{figura beta m}
\end{figure}

We can also obtain an alternative estimation of $\beta$ based on the actual observational upper value of frequency. If we assume that for a given NS $\epsilon$ is only a fraction $\beta_{o}$ of $\epsilon_{max}$, i.e.
\begin{equation}
    \epsilon=\beta_{o}\epsilon_{max};\,\,\,\,\beta \leq 1,
    \label{definizione beta}
\end{equation}
%
one can express the equilibrium frequency as a function of $\beta_{o}$, as in Eq \eqref{nu equilibrio epsilon max}, namely
\begin{equation}
    \nu_{eq}(\beta_{o})= C\frac{\dot{M}^{1/9}(Ma)^{1/18}}{I^{2/9}\beta_{o}^{2/9}\Tilde{\epsilon}^{2/9}},
    \label{nu di beta}.
\end{equation}
Then, given an EoS and fixed the stellar mass, we state that the minimum feasible value of $\beta_{o}$ is the one that satisfies the condition
\begin{equation}
    \nu_{eq}(\beta_{o})=\nu_{o}.
    \label{beta di m}
\end{equation}
%
Using this selection criterion for different masses $M$ and different mass accretion rates $\dot{M}$ we construct the curves shown in Fig. \ref{figura beta m}. 
In all the cases $\beta_{o}<1$, and for the smaller mass accretion rate $\dot{M}=10^{-10}M_{\odot}/\mathrm{yr}$ (with $M=1.4M_{\odot}$), we find, for the SLy EoS, $\beta_{o}=0.002$, while $\beta_{o}=0.0007$ for the stiffer BSk21. 

\subsection{Millisecond pulsars and gravitar limit}

\begin{figure}
\includegraphics[width=\columnwidth]{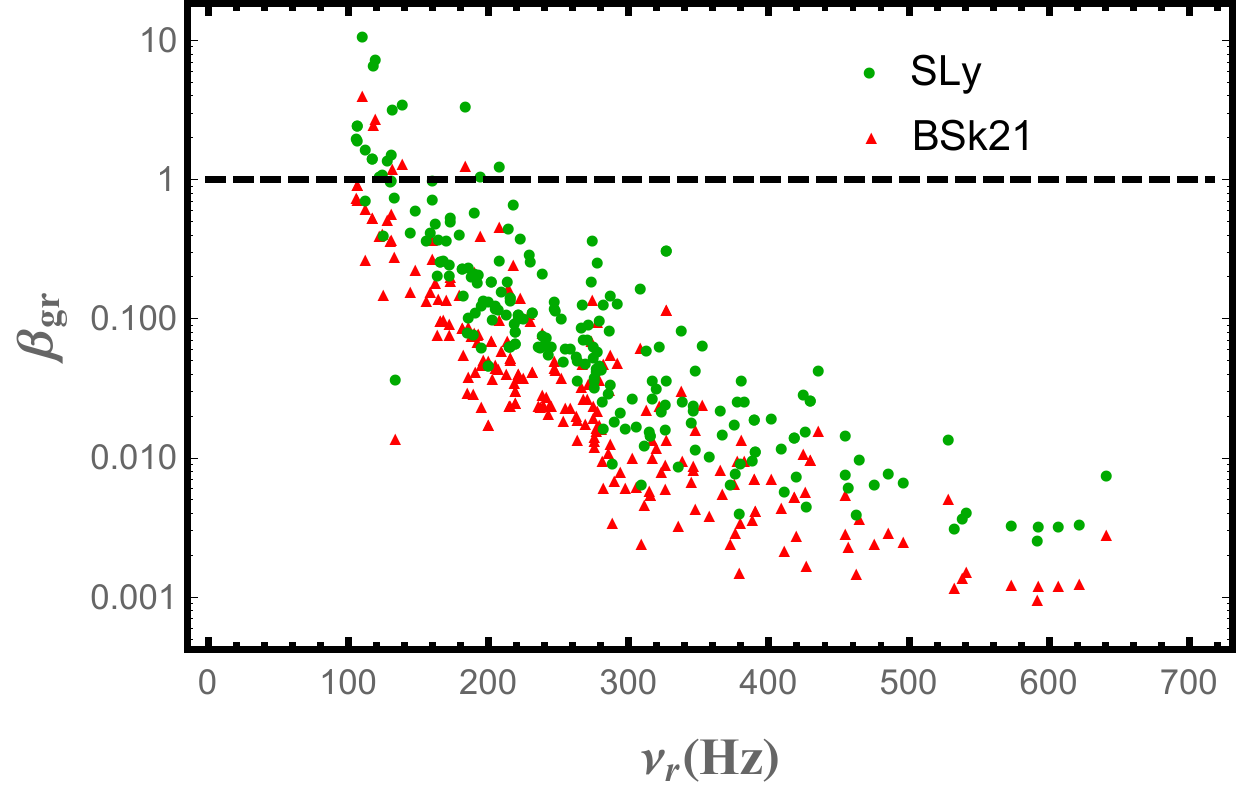}
\caption{Ratio of the observational ellipticity in the gravitar limit $\epsilon_{gr}$ (Eq \eqref{epsilon grav}) and our upper limit $\epsilon_{max}$ for a $M=1.4M_{\odot}$ NS with $\gamma^*=2$. Green dots are calculated with SLy EoS while red triangles with BSk21. The dashed, black line indicates $\beta_{gr}=1$.}
\label{beta di nu}
\end{figure}

In the previous two sub-sections we focused only on accreting NSs, while in the following the analysis is extended also to spinning-down objects. 
Millisecond pulsars are thought to be the evolutionary descendant of accreting objects \citep{Alparmilli,Bhattamillisecond}, i.e. old NSs that have been spun up to high rotational frequencies via accretion. If we assume that LMXB objects can develop large ellipticity due to starquakes, we can expect that millisecond stars too have a non-zero $\epsilon$, i.e. a residual part of their initially larger quadrupolar deformation.\footnote{For these objects it could be useful to take into accounts also the long-time evolution of elastic layers due to a non-null viscosity. However, at the present time, the viscosity $\nu$ of NSs' crust is essentially unknown, even if in the last years some first estimations have appeared \citep{Kwang,samviscoso}. The inclusion also of this parameter in a consistent model can help the understanding of the global dynamics of a realistic NS, and is clearly a very interesting field for future research.} In this case we can introduce a simple model to explain the millisecond actual decreasing period. In fact, these objects lose energy via electromagnetic and GWs emission. Therefore, following \citet{woan2018}, that assume $I=10^{45}\,\mathrm{g\,cm^2}$ and a vacuum dipole radiation, we can write
\begin{align}
    \frac{\dot{P}}{10^{-20}}=0.98\left(\frac{1\mathrm{ms}}{P}\right)\left(\frac{I_0}{10^{45}\mathrm{g\,cm^2}}\right)^{-1}\left(\frac{B_{s}}{10^{8}G}\right)+\\ 
    +2.7\left(\frac{1\mathrm{ms}}{P}\right)^{3}\left(\frac{I_0}{10^{45}\mathrm{g\, cm^2}}\right)\left(\frac{\epsilon}{10^{-9}}\right)^{2},\nonumber
    \label{pdot bryn}
\end{align}
where $B_{s}$ is the surface magnetic field of the star.
In order to to calculate the ellipticity needed to explain the observed stellar spin-down in the case of pure gravitational wave emission, i.e. in the \emph{gravitar limit} \citep{palomba2005}, we can neglect the magnetic field and put $B_{s}=0$ in Eq (\textcolor{blue}{35}). Solving for $\epsilon$ (the subscript \emph{gr} stands for gravitar) we get
%
\begin{equation}
    \epsilon_{gr}=10^{-9}\sqrt{0.37 \left(\frac{I_0}{10^{45}\mathrm{g \,cm^{2}}}\right)\left(\frac{\dot{P}}{10^{-20}}\right)\left(\frac{P}{1\mathrm{ms}}\right)^{3}}.
    \label{epsilon grav}
\end{equation}
We can compare this value with our upper limit, Eq \eqref{epsilon max}, assuming a standard star configuration (with $M=1.4M_{\odot}$ and an adiabatic index $\gamma^*=2$). 
As representatives of millisecond pulsars we select all the non-accreting stars with $\nu_{r}>100$ Hz in the ATNF database  (\url{http://www.atnf.csiro.au/research/pulsar/psrcat/}).
In Fig. \ref{beta di nu} we show the ratio
\begin{equation}
\beta_{gr}=\frac{\epsilon_{gr}}{\epsilon_{max}}.    
\end{equation}
The spin-down of all the stars with $\beta_{gr}\leq1$ can be explained with pure gravitational wave emission due to starquakes, since the ellipticity necessary to produce the current stellar spin-down is smaller than the value of $\epsilon_{max}$ corresponding to the NS's rotational frequency.

On the contrary, we observe that for all the objects with $\beta_{gr}>1$ we need to invoke a non-null magnetic field to explain the current observed spin-down, i.e. to abandon the gravitar hypothesis, using the full expression of Eq (\textcolor{blue}{35}).
The minimum value of $\beta_{gr}$ in our sample is 
\begin{equation}
    \beta_{gr}=0.001.
    \label{valore beta 1.4}
\end{equation}
%
In the case of SLy roughly $90\%$ of the selected star have $\beta_{gr}<1$, while for the stiff BSk21 EoS the percentage rise to $97\%$.
These results show that the starquakes mechanism can produce very large ellipticity, and thus that actual value of $\epsilon$ for millisecond pulsars can in principle be produced by crust rupture on their progenitors accreting stars.
%
Fig \ref{beta di nu} shows also another interesting aspect. The slowest NSs in our catalogue have a $\beta_{gr}$ value that is larger than $1$, which means that for these objects our upper limit is smaller than the gravitar ellipticity value.

Finally we note that also the recent estimation of J1023+0038 pulsar's ellipticity \citep{Bhatta2020} confirm the range of expected $\beta$ form LMXBs. In fact, for a $1.71\pm0.16M_{\odot}$ NS with $\gamma^*=2$, the ratio between Bhattacharyya's ellipticity and our maximum values is about $10^{-3}$ both for SLy and BSk21 EoSs.


\section{Conclusion}
GWs emission from rapidly rotating NSs is a very actual and interesting field of research, and so the study of the maximum mountains on pulsars. 
Many scenarios as been invoked to produce a non-null ellipticity, such as thermal mountains or particular magnetic fields configurations; other studies focused on the maximum quadrupole moment that a NS can sustain.
In this paper it is presented for the first time (at least at the best of our knowledge) a realistic and consistent calculation of a new mechanism to produce GWs emission: starquakes.
This kind of mechanism has been already recently proposed by \citet{fattoyev2018}, but here it is the first time that the full problem is consistently studied. In fact, we construct a model to study the reaching of crust breaking, the maximum ellipticity produced in a sequence of crust ruptures and the equilibrium frequency reached by accreting object.
Our calculations shows that NSs crust can fail due to centrifugal stresses when the frequency of the rotating star is in the range $200-900$ Hz, depending on the EoS and the mass of the star. 
In general, the equilibrium frequency is found to be smaller than the breaking one and, therefore, it is also below the actual observable threshold of $716.36$ Hz.
The study of how large the ellipticity due to starquakes can be has given an upper limit for $\epsilon$ lying between $10^{-9}$ and $10^{-5}$, depending on the EoS and the mass of the object. The comparison between different EoS showed that the stiffer ones produce larger ellipticities and, consequently, bring the star towards a lower equilibrium frequency.

In the case of highly spinning pulsars, and for $\gamma$ different from $\gamma^*$, we also showed that crust failure can produce an ellipticity comparable with the maximum theoretically expected value given by \citet{owen2013}. But with a great difference; in fact, the mechanism producing the quadrupolar deformation is due to crust breaking in our model, instead of be sustained by the elastic crust.

We  found that the stellar mass affects the star's response and that a $2 M_{\odot}$ objects creates  an ellipticity about one order of magnitude smaller than the lighter $1 M_{\odot}$ ones.
Moreover, $\epsilon$ depends strongly also on the adiabatic index: even just a small difference from the adiabatic equilibrium
value leads to large differences on the ellipticity.

Last, but not least, comparing our upper limit estimation with the observational data coming from both accreting objects and millisecond
pulsars, and calculating within our model the ellipticity  deduced from the data, we found a value of $\epsilon$ that is even a very small fraction of our model upper limit ($\epsilon/\epsilon_{max} =10^{-3}$) can in principle explain the vast majority of the available data ($90\%$ to $97\%$ depending on the EoS used). 

Our model explains why rotating NSs with frequencies greater than about 700 Hz are not observed and, at the same time, it expects that accreting NSs stars could generate, with the right combination of distance, mass and frequency, GWs of intensity detectable by LIGO-Virgo O3 run ,which complete data analysis is still ongoing.

It also shows that the evolutionary scenario depicted in this paper seems to be sufficiently  robust to be a competitive candidate in describing NSs achieve a  dynamical equilibrium.

All these results are obtained in a Newtonian framework. A natural and important step for further improvement of this model is its generalisation to General Relativity, which permits also to use realistic EoSs for the description of the whole star.

\section*{Acknowledgments}
The authors are grateful for useful discussions with L. Perotti, B. Haskell, M. Antonelli and P. M. Pizzochero.
\section*{Data availability}
The data underlying this article will be shared on reasonable request to the corresponding author.


\bibliographystyle{mnras}
\bibliography{biblio_articolomodi}


\appendix
\section{Inertia and Spherical Harmonics Expansion}\label{appendice a}
The inertia tensor is defined as 
\begin{equation}
I_{ij}=\int\rho\left(r\right)\left(r^{2}\delta_{ij}-r_{i}r_{j}\right)dV.
\label{eq definizione inerzia}
\end{equation}
The initial, non-rotating stellar configuration will be deformed by centrifugal force, that gives rise to a change of the star's density profile, can be written as
\begin{equation}
\rho\left(r\right)=\rho_{0}\left(r\right)+\rho^{\Delta}\left(r,\theta,\phi \right),
\end{equation}
where we have put in evidence the initial, unstressed profile $\rho_{0}$ and the local perturbation $\rho^{\Delta}$. 
We can use the spherical symmetry of the problem to recast Eq \eqref{eq definizione inerzia}. To that purpose let us first introduce the spherical harmonics, defined as 
\begin{equation}
Y_{\ell m}\left(\theta,\varphi\right)=P_{\ell m}\cos\left(\theta\right)e^{im\varphi},
\end{equation}
where $P_{\ell m}$ are the associated Legendre polynomials
\begin{equation}
P_{\ell m}\left(x\right)=\begin{cases}
\frac{1}{2^{\ell}\ell!}\left(1-x^{2}\right)^{m/2}\frac{d^{\ell+m}\left(x^{2}-1\right)^{\ell}}{dx^{\ell+m}} & m>0\\
\left(-1\right)^{m}\frac{\left(\ell-m\right)!}{\left(\ell+m\right)!}P_{\ell m}\left(x\right) & m<0
\end{cases}
\end{equation}
Now, by using the spherical harmonics expansion we can write the star's density perturbation as
\begin{equation}
\rho^{\Delta}\left(r,\theta,\varphi\right)=\sum_{\ell=0}^{\infty}\sum_{m=-\ell}^{m=\ell}\rho_{\ell m}^{\Delta}\left(r\right)Y_{\ell m}\left(\theta,\varphi\right)
\end{equation}
and the total perturbed potential as
\begin{equation}
    \Phi^{\Delta}\left(r,\theta,\varphi\right)=\sum_{\ell=0}^{\infty}\sum_{m=-\ell}^{\ell}\Phi_{\ell m}^{\Delta}\left(r\right)Y_{\ell m}\left(\theta,\varphi\right).
\end{equation}
In this work, all the perturbed terms are the ones due to rotation, thus we can focus only on the centrifugal potential $\phi^{C}$. It can be expanded as a sum of the only $\ell=0$ and $\ell=2$ terms, i.e.
\begin{equation}
\phi^{C}\left(r,\theta,\varphi\right)=\phi_{00}^{C}\left(r\right)Y_{00}\left(\theta,\varphi\right)+\sum_{m=-\ell}^{m=\ell}\phi_{2m}^{C}\left(r\right)Y_{2m}\left(\theta,\varphi\right).
\end{equation}
If we substitute the density expansion in Eq \eqref{eq definizione inerzia}, we can express the inertia tensor $I$ as
\begin{equation}
I=I_{0}+I^{\Delta},
\end{equation}
where we highlighted the unperturbed tensor of inertia $I_0$. With some straightforward algebra the perturbed inertia tensor can be divided into two terms
\begin{equation}
I^{\Delta}=I_{00}^{\Delta}+I_{20}^{\Delta}.
\end{equation}
Note that by choosing a coordinate system in which the rotational axis $z$ coincides with the one at $\theta=0$, the centrifugal potential contains only the $m=0$ order of the $\ell=2$ harmonic term. 
The contributions of these spherical harmonic terms perturbed to the inertia tensor are, respectively, 
\begin{equation}
I_{00}^{\Delta}=\frac{8\pi}{3}\delta_{ij}\int_{0}^{a}\rho_{00}^{\Delta}\left(r\right)r^{4}dr,
\end{equation}
and 
\begin{equation}
I_{20}^{\Delta}=\left(\frac{1}{3}\delta_{ij}-\hat{r}_{i}\hat{r}_{j}\right)\frac{4\pi}{5}\int_{0}^{a}\rho_{20}^{\Delta}\left(r\right)r^{4}dr.
\end{equation}
For simplicity, in the main text we use the notation
\begin{equation}
\Delta I=I_{20}^{\Delta}.
\end{equation}

\section{ALTERNATIVE ESTIMATION OF ELLIPTICITY}
\begin{figure}
\includegraphics[width=\columnwidth]{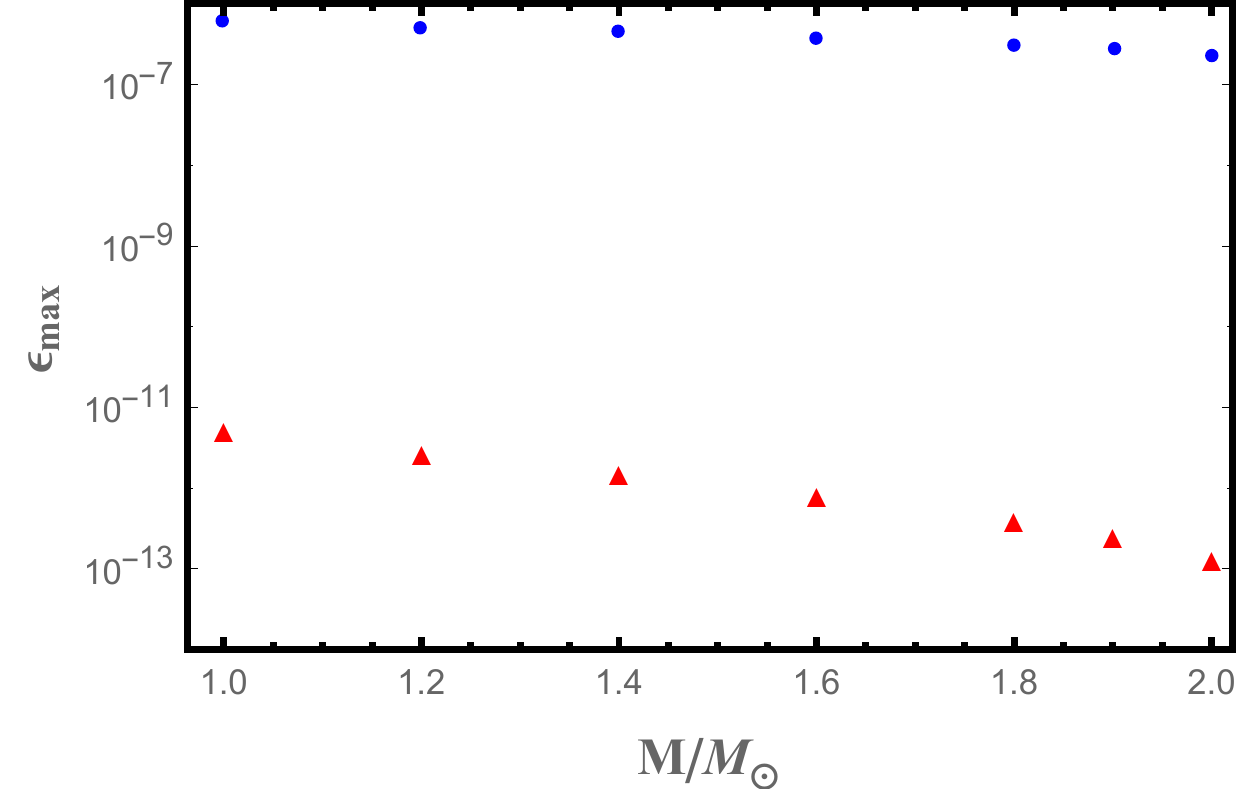}
\caption{Maximum elliptiticy given by a sequence of starquakes, calculated as the difference between the fluid and the deformed elastic configuration as in Eq \eqref{epsilon max}, showed as blue dots, and as the difference between the fluid and the spherical elastic configuration, Eq \eqref{epsilon fluido max}, reported as red triangles. The ellipticity is calculated for different stellar masses ranging from $1 M_{\odot}$ to $2M_{\odot}$, keeping $\nu=1$ Hz and $\gamma^*=2$ fixed.}
\label{confronto elliticita non rot rot}
\end{figure}

Another heuristic estimation of ellipticity can be obtained by assuming that the elastic crust can keep the NS in a spherical configuration, despite the centrifugal forces due to rotation \citep{fattoyev2018}. In this case $\Phi_{20}^{\Delta E}\left(a\right)=0$ and one gets (see Eq \ref{epsilon max})
\begin{equation}
    \epsilon_{max}=\frac{a^{3}}{3I_{0}G}\Phi_{20}^{\Delta F}\left(a\right).
\label{epsilon fluido max}
\end{equation}
We observe this rough approximation to be larger than our $\epsilon_{max}$, since the rotating configuration is squeezed towards the equatorial plane by the fast rotation. In Fig \ref{confronto elliticita non rot rot} we compare, for different NSs masses and $\nu=1$ Hz, the ellipticity given by our upper limit $\epsilon_{max}$ and the one given by Eq \eqref{epsilon fluido max} for the case $\gamma^*=2$. The first of the two ellipticity values has clearly a different dependence on the stellar mass and, furthermore, it is 5 orders of magnitude smaller than the second. 

\bsp	
\label{lastpage}
\end{document}